# Two-Photon Absorption by $H_2$ Molecules: Origin of the 2175Å Astronomical Band?

Peter P. Sorokin[1*] and James H. Glownia[2]

**The near UV spectra of OB stars in our galaxy are often dominated by a very broad extinction band peaking at ≈2175Å. Forty years after its discovery, the origin of this band remains unknown, although interstellar dust particles are generally assumed to be the carriers. Here we report that two-photon absorption by $H_2$ molecules in gaseous clouds enveloping OB stars can lead to a band peaking at ≈2175 Å. We present astronomical spectral evidence supporting our proposal that this nonlinear absorption mechanism accounts for the λ2175 feature.**

[1] IBM Research Division, T.J. Watson Research Center, Yorktown Heights, NY 10598-0218, USA. [2]MST-CINT, MS K771, Los Alamos National Laboratory, Los Alamos, NM 87545-1663, USA.

[*]To whom correspondence should be addressed. E-mail: sorokin@us.ibm.com

The near UV spectra of OB stars in our galaxy are often dominated by a very broad extinction band peaking at ≈2175Å. Forty years after its discovery, the origin of this band remains unknown, although the general consensus of astronomers is that the band results from linear absorption or linear scattering by interstellar dust particles. A recent article published in this journal (*1*) summarizes mainstream thinking regarding the origin of the mysterious λ2175 astronomical extinction band.

Recently, the present authors have proposed an alternative explanation for the λ2175 extinction band whereby the latter is produced via a nonlinear photonic mechanism involving two-photon absorption by $H_2$ molecules present in the gases that immediately surround OB-type stars (*2*). At first glance, this proposal might seem to be rather implausible, since two-photon absorption events normally occur with high probability in any medium only if the latter is simultaneously irradiated by intense, ideally monochromatic, light - such as that produced by a laser. Up to now, the possibility that intense UV or visible coherent light could somehow be generated in the vicinities of stars has not been seriously considered, although the existence of microwave and far infrared masers in Space is by now very well established. However, we recently developed a physically realistic photonic model which can explain how intense laser-like light in the vacuum ultraviolet (VUV) can be generated in gases surrounding OB-type stars (*2,3*). In this model, intense, monochromatic, coherent light at the resonance frequencies of select ionic and/or atomic and/or molecular species that are dominantly present in the gases immediately surrounding such stars is generated via the process of stimulated Rayleigh scattering. Due to the constraints imposed by conservation of both (linear) momentum and energy (*2,3*), the intense coherent light that is generated in this nonlinear process must always propagate radially inwards towards the photosphere of the illuminating star in a spherically symmetrical fashion. Thus, this radiation cannot in principle be directly discerned by an observer looking along the line-of-sight to the star! What *can* easily be discerned by the observer is the pump light that is used to drive the stimulated Rayleigh scattering process. The pump light is the incoherent continuum light emitted outwardly from the photosphere of the OB star in the spectral vicinities of the resonance lines of the generating species. As with any stimulated scattering process (*e.g.* stimulated Raman scattering), stimulated Rayleigh scattering is characterized by a pump power threshold, below which the process entirely ceases. An observer can easily note whether or not the stimulated Rayleigh scattering threshold has been exceeded in a given star by examining whether or not a wide spectral interval of continuum light adjacent to the resonance line of the generating species is "missing" (*2,3*). The authors suggest that the sequence of spectra presented in Fig. 2 of (*3*) provides a particularly good illustration of the dramatic spectral changes that can occur in OB stars when the threshold for stimulated Rayleigh scattering is crossed.

In (*3*), the basic idea that stimulated Rayleigh scattering could occur in the immediate vicinities of bright stars was initially conceived, and the main physical attributes of this nonlinear photonic mechanism were deduced. The model was then exclusively applied to account for selective line-driven acceleration of ions in the stellar winds of OB-type stars.

In (*2*), we extended the model of (*3*) to show that nonlinear photoexcitation of $H_2$ molecules could also theoretically occur via stimulated Rayleigh scattering when sufficient densities of the former are present in the gases that immediately surround OB stars. The present report focuses on a specific idea suggested in (*2*), namely, that the λ2175 astronomical extinction band originates from two-photon absorption (TPA) by $H_2$ molecules in gaseous clouds enveloping OB stars. Here we utilize archival astronomical spectral data to show that the observed strength of the λ2175 band in any OB star strongly correlates with the spectral evidence that nonlinear $H_2$ photoexcitation is occurring in the star, in effect supporting the basic ideas mentioned above (a listing of some such stars is contained in the supporting online materials).

In the TPA model for the λ2175 band, it is assumed that intense, monochromatic, coherent light is continuously being generated via stimulated Rayleigh scattering at the frequencies of $H_2$ B-state and C-state resonance lines originating from levels J"=0 and J"=1 of X0, the lowest vibrational state of the ground electronic state X. [Less intense coherent light is also sometimes generated at the frequencies of resonance lines originating from (X0, J"=2) and (X0, J"=3).] This coherently generated light continually propagates in a spherically symmetric manner radially inwards towards the illuminating OB-type star, effectively providing intense "first step" monochromatic radiation needed for $H_2$ two-photon transitions to occur. (Despite our use here of the term "first step", absorption of the two photons in each TPA event should be viewed as occurring simultaneously, not sequentially.)

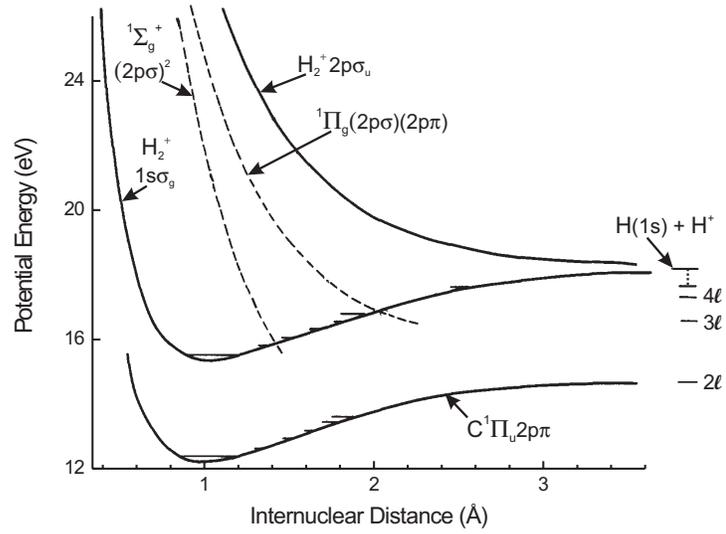

**Fig. 1.** Potential curves for some excited states of $H_2$ and for two states of the ion $H_2^+$. Doubly excited states of $H_2$ are shown as dashed lines. Figure adapted from Fig. 2 of (*6*).

Accepting the validity of the above assumption, one can then immediately see from a partial $H_2$ energy level diagram (Fig. 1) that continual absorption of continuum light emitted from the star's photosphere should in principle occur as "second steps" in resonantly-enhanced two-photon transitions. (TPA allows absorption of "second step" light to occur, even when such light and the "first step" light are counterpropagating.) The frequencies of these "second step" transitions should match the (vertical) energy separations between the doubly excited $^1\Pi_g(2p\sigma)(2p\pi)$ state and various B- or C-state levels, or between the doubly excited $^1\Sigma_g^+(2p\sigma)^2$ state and B-state levels. These two-photon transitions are fully allowed, since the electronic configurations of the X state, the B state, and the C state are, respectively, $^1\Sigma_g^+(1s\sigma)^2$, $^1\Sigma_u^+(2p\sigma)(1s\sigma)$, and

$^1\Pi_u(2p\pi)(1s\sigma)$. From Fig. 1, one can see that "second step" transitions originating from B or C states would, in general, have very broad bandwidths, since the vertical projections from a given bound lower level to either unbound upper level would span a very large frequency range, due to the steepness of the doubly excited state potential curves. Transitions originating from B0 or C0 would have the narrowest bandwidths, since the $H_2$ internuclear distance varies much less in these states. From Fig. 1, one can see that the "second step" transition originating from C0 would be centered in the far UV − at a wavelength less than 1000 Å. The B-state potential curve is not shown in Fig. 1. However, it is known (*4*) that the B0 levels occur at about 11.18 eV, and that the wavefunctions for these levels have maximum probabilities of occurrence at an $H_2$ internuclear distance of about 1.3 Å. From Fig.1, one sees that the doubly excited $^1\Sigma_g^+$ state appears to have an energy of about 17 eV at an $H_2$ internuclear separation of 1.3 Å. Therefore, the "second step" transition involving B0 as resonant intermediate state should roughly peak at $\approx 2130$ Å, which lies suggestively close to the λ2175 band maximum. Predicted "second step" center wavelengths corresponding to the vertical separations between the doubly excited $^1\Pi_g$ state and various B- and C-state vibrational levels are indicated in Fig. 4 of (*2*) (also reproduced in supporting online materials). As shown in that figure, these predicted wavelengths all lie within the wavelength region spanned by the spectral profile of the λ2175 feature.

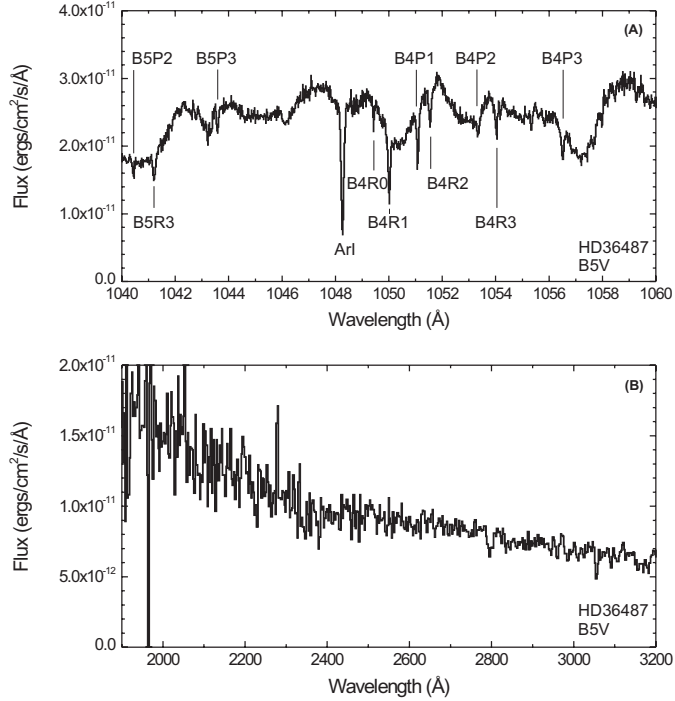

**Fig. 2.** Spectra of the B5V star HD 36487 recorded in different spectral ranges. **(A)** The portion of the *FUSE* spectrum between 1040Å and 1060Å. **(B)** The portion of the *IUE* (low dispersion) spectrum between 1900Å and 3200Å.

If our proposed TPA explanation for the λ2175 extinction band were indeed correct, one should be able to observe a strong correlation between the strength of the λ2175 band in the line-of-sight to any OB-type star and some measure of *nonlinearly excited* $H_2$ in the same line-of-sight. To explore whether or not such a correlation exists, we have utilized the *Multimission Archive at STScI (MAST)*, which is easily accessed through the *MAST Scrapbook* interactive web site. Using this resource, we have examined the spectra of approximately 100 OB-type stars for which the *MAST Scrapbook* web site contains both spectra that span the wavelength region of the λ2175 extinction band [mostly spectra recorded by the *International Ultraviolet Explorer (IUE)* orbiting satellite, a few by the *Wisconsin Ultraviolet Photo-Polarimeter Experiment (WUPPE)*] as well as spectra that span the $H_2$ $(X0 \rightarrow B, C)$ absorption regions [mostly spectra recorded by the *Far Ultraviolet Spectroscopy Explorer (FUSE)*, a few by the *Tubingen Ultraviolet Echelle Spectrometer (TUES)*]. From this examination, we made the following observations.

(A) If there is *no* apparent $H_2$ VUV absorption in a given line-of-sight, or if one observes only relatively weak, non-saturated, absorption lines at the $H_2$ $(X0 \rightarrow B, C)$ frequencies (*e.g.* as in Fig. 2A), with lines originating from (X0, J"=0 and 1) *not* being orders-of-magnitude more absorbing than lines originating from higher-J" levels (thus suggesting the occurrence of *linear* absorption by $H_2$ molecules present at a relatively high effective temperature in the cloud of gas surrounding the star), then the λ2175 band will invariably be *undetectable* in that same line-of-sight (Fig. 2B). Either of these two situations can sometimes prevail in lines-of-sight to surprisingly hot stars (Figs. 3A, 3B, and 3C).

(B) On the other hand, if the λ2175 absorption feature is very prominently seen in the line-of-sight to an OB star (*e.g.* as in Fig. 4B or 4C), then that same line-of-sight will also invariably contain spectral evidence strongly suggesting the occurrence of $H_2$ stimulated Rayleigh scattering in the gas cloud surrounding the star under observation. That evidence consists of an $H_2$ VUV spectrum in which continuum light from the star at wavelengths corresponding to allowed $H_2$ transitions originating from (X0, J"=0 and 1) is absorbed *orders-of-magnitude* more strongly than light at wavelengths of allowed transitions originating from higher rotational levels of X0 (Fig. 4A). Most of the roughly 100 OB stars examined in the course of the present study (*5*) fall under category (B).

On the basis of the $H_2$ nonlinear photonic mechanism here being proposed, it is apparent that one can reasonably account for the main aspects of recorded astronomical spectra that involve the λ2175 extinction band. Basically, one first postulates that a cloud of hydrogen gas closely envelops an OB star. If the $H_2$ densities in the enveloping gas cloud are too low, or if the continuum radiation emitted by the OB star is too weak, there will be insufficient optical gain for the threshold of stimulated Rayleigh scattering to be exceeded. In this case, only randomly occurring ($H_2$ photodissociation / H-atom recombination) events will take place in the gas cloud. As a result of these events, a characteristic steady-state distribution of the $H_2$ population among the various lower rotational levels of X0 will be produced, resulting in the appearance of $H_2$ linear absorption spectra such as those shown in Figs. 2A and 3A. No λ2175 extinction band will be present in the line-of-sight. However, if within the gaseous cloud that envelops an OB star, the product of the $H_2$ density in either (X0, J"=0) or (X0, J"=1) times the stellar continuum flux per unit frequency width in the spectral vicinities of strongly allowed $H_2$ transitions originating from these levels exceeds a certain critical value, then the threshold for stimulated Rayleigh scattering can become exceeded. When this happens, (i) anomalously strong, broadband absorption of continuum light spectrally located on either side of the (R0, R1, and P1) $X0 \rightarrow B$ and (R0, R1, and Q1) $X0 \rightarrow C$ transitions will occur, with such absorption being directly observable in the line-of-sight; (ii) intense monochromatic coherent VUV light at the above $H_2$ transition frequencies that propagates radially inwards towards the photosphere of the star in a spherically symmetrical manner will be generated, but will not be observable in the line-of-sight; and (iii) a strong λ2175 extinction band resulting from two-photon absorption by $H_2$ molecules present in the cloud enveloping the OB star will be clearly apparent in the line-of-sight.

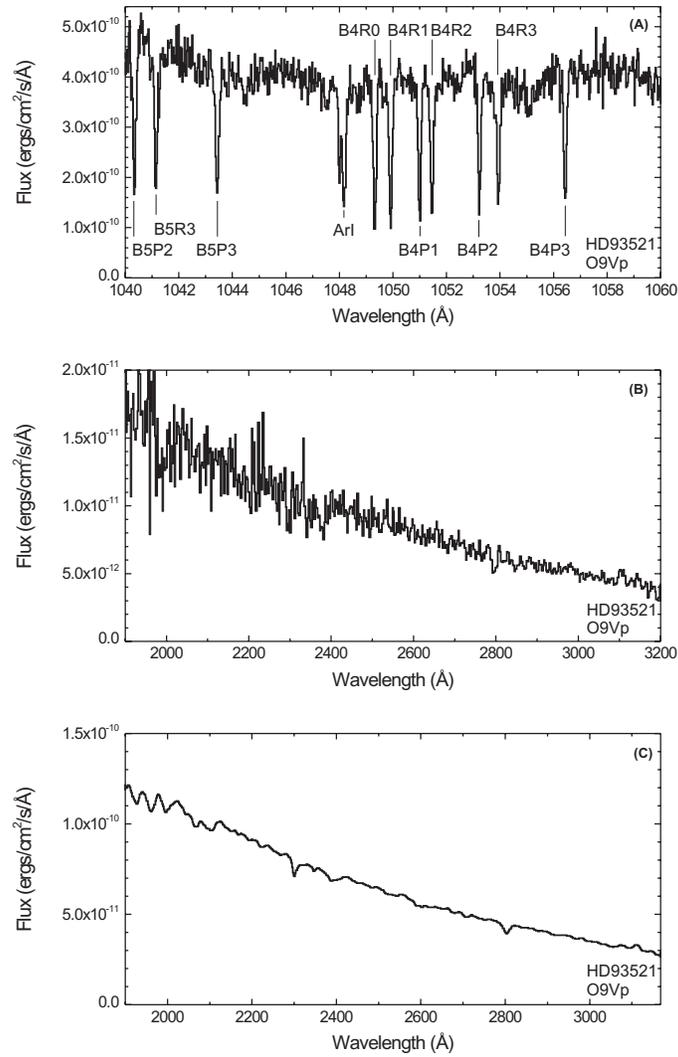

**Fig. 3.** Spectra of the O9Vp star HD 93521 recorded in different spectral ranges. **(A)** The portion of the *TUES* spectrum between 1040Å and 1060Å. **(B)** The portion of the *IUE* (low dispersion) spectrum between 1900Å and 3200Å. **(C)** The portion of the *WUPPE* spectrum between 1900Å and 3168Å.

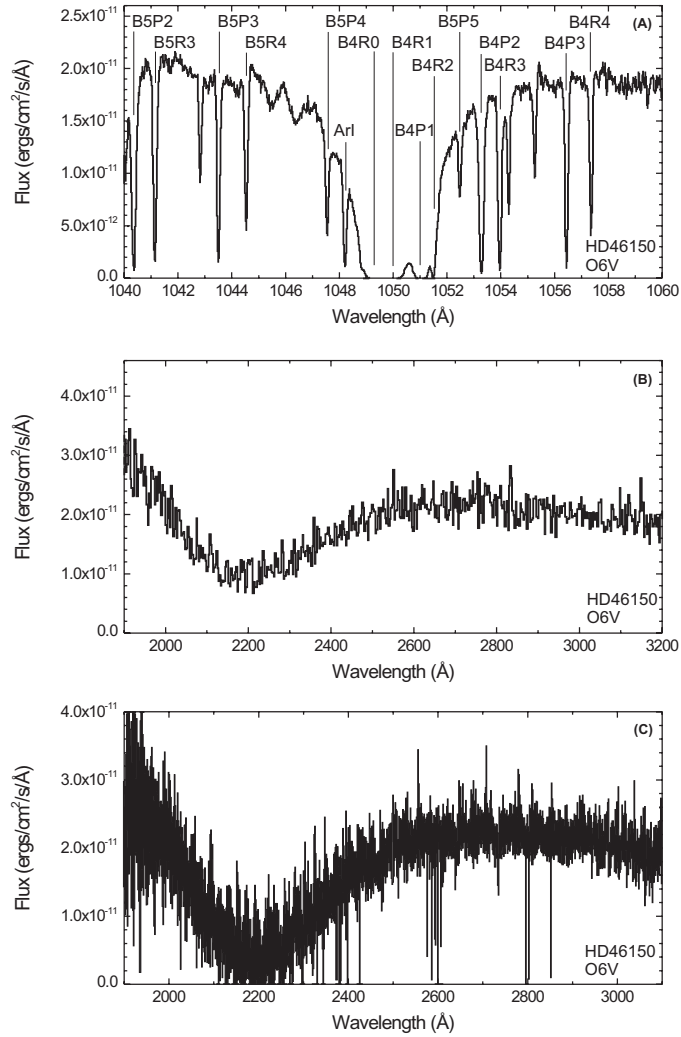

**Fig. 4.** Spectra of the O6V star HD 46150 recorded in different spectral ranges. **(A)** The portion of the *FUSE* spectrum between 1040Å and 1060Å. **(B)** The portion of the *IUE* (low dispersion) spectrum between 1900Å and 3200Å. **(C)** The portion of the *IUE* (high dispersion) spectrum between 1900Å and 3100Å.

7. P.P.S. is grateful to IBM Research management for having allowed him retirement use of office and library facilities for the past five years. J.H.G. acknowledges partial support for this work through the Los Alamos National Lab LDRD-ER program.

**Supporting online materials:**

For convenience, we reproduce Fig. 4 of (*2*). In this figure, the predicted "second step" center wavelengths corresponding to the vertical separations between the doubly excited $^1\Pi_g$ state and various B- and C-state vibrational levels are indicated. It is seen that the predicted wavelengths match the spectral profile of the λ2175 feature. Here we also provide a table listing some OB-type stars examined where we can directly correlate a signature of nonlinearly excited $H_2$ with the strength of the λ2175 band (along common lines-of-sight). In creating this table, we utilized archival astronomical spectral data.

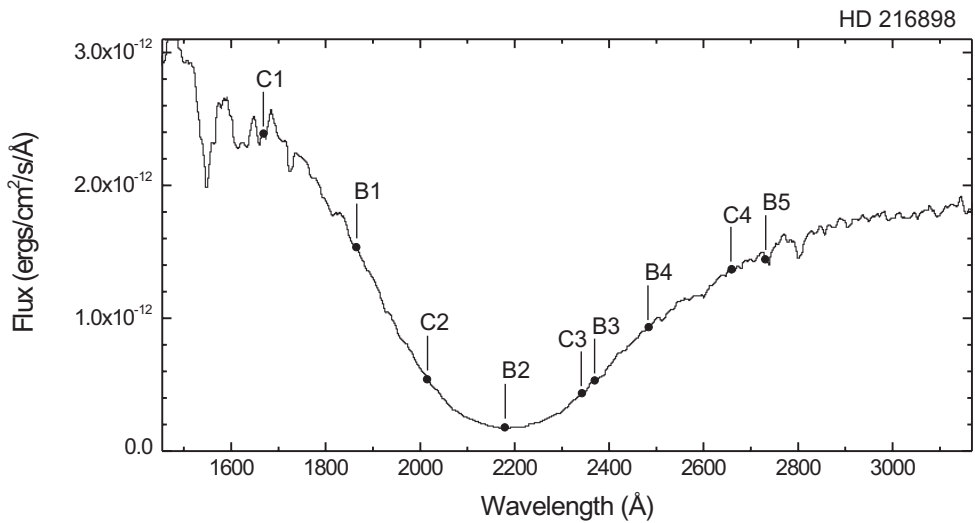

Fig. 1. WUPPE spectrum of HD 216898, downloaded from the MAST Scrapbook interactive website. Predicted spectral locations of two-photon absorption band centers are superimposed (see main text).

Table 1. Listing of the Strong Correlation Observed in 105 OB Stars examined between the Nonlinear Absorption Intensity of $H_2$ and the Strength of the λ2175Å Astronomical Extinction Feature (thus far, without exception, all stars examined support our model):

**#1   HD 37373**   Coordinates are 05 37 50.83; -06 43 19.5. This star has no λ2175 absorption evident on the *IUE l.d.* (low-dispersion) spectrum. (For all the stars in this table, we viewed the *IUE l.d.* spectrum between 1900Å and 3200Å.) The *FUSE* spectrum between 1040Å and 1060Å shows only the A I absorption line at 1048.22Å, and one $H_2$ absorption line – B4-0R1 (hereafter abbreviated to B4R1). The B4R1 line appears (very weakly) close to its calculated wavelength of 1049.96Å. The broad modulations of the star's continuum seen in the *FUSE* spectrum represent stellar photospheric absorptions.

**#2   HD 35580** (B8.5V) Coordinates are 05 22 22.15; -56 08 3.8. This star has no λ2175 absorption evident on the *IUE l.d.* spectrum. In the 1040-1060Å *FUSE* spectrum, there is only a very slight hint of the A I absorption line at 1048.22Å, but *no* lines at all due to $H_2$. The wide stellar continuum modulations seen are almost identical to those in HD 37373.

**#3   HD 36487** (B5V) Coordinates are 05 31 41.44; -07 02 55.1. This star has no λ2175 absorption evident on the *IUE l.d.* spectrum. The 1040-1060Å *FUSE* spectrum of HD 36487 has broad stellar photospheric absorptions appearing very similar to those in HD 37373 and HD 35580. However, HD 36487 absorbs more strongly than HD 37373 at $H_2$ wavelengths. In HD 36487 the line at B4R1 is stronger, and there are also weakly present absorption lines at B4R0, B4P1, B4R2, B4P2, B4R3, B4P3, B5P2, B5R3, and B5P3. The A I line at 1048.22Å is present about as strongly as in star #1. See Fig. 2 in the manuscript.

**#4   HD 37525 (**B5V) Coordinates are 05 39 1.49; -02 38 56.4. This star has no λ2175 absorption evident on the *IUE l.d.* spectrum. The 1040-1060Å *FUSE* spectrum is similar to that of star #3, except that the $H_2$ lines are here two-to-three times stronger. The modulations of the stellar continuum are generally similar to those appearing in stars (#1-3). The A I line here is as strong as in stars #1 and #3.

**#5   HD 37526** (B3V) Coordinates are 05 39 2.40; -05, 11, 40.1. This star has no λ2175 absorption evident on the *IUE l.d.* spectrum. The $H_2$ absorption lines in the 1040-1060Å *FUSE* spectrum have roughly the same relative intensities as they have in star #4, only here the strengths of the lines are about fifty percent weaker. The A I line at 1048.2Å is here about as strong as in star #4. The stellar continuum modulations are here generally similar appearing to those in stars #1- #4.

**#6  HD 93521** (O9Vp) Coordinates are 10 48 23.51; +37 34 13.1. This star has no λ2175 absorption evident on the *IUE l.d.* spectrum, and also on the 1900Å-3168Å (very low noise) *WUPPE* spectrum. No *FUSE* spectrum is available for this star, but there exists a *TUES* spectrum. In the spectral region 1040Å-1060Å, the *TUES* spectrum displays roughly equally intense $H_2$ absorption lines at B4R0, B4R1, B4P1, B4R2, B4P2, B4R3, and B4P3. The $H_2$ absorption lines occurring at B5P2, B5R3, and B5P3 also fall in the 1040Å- 1060Å spectral region. These are observed to have roughly the same intensities as the above mentioned B4-0 lines. However, the absorption intensities of both B5R4 at 1044.54Å and B5P4 at 1047.55Å are observed to be at least ten times less than those of B5-0 transitions originating from X0 levels with J" equal to or less than 3. The same large drop of absorption intensity is observed to occur for B4-0 transitions originating from X0 levels higher than X0, J''=3. See Fig. 3 in the manuscript.

---

**#7  HD 100340**  (B1V) Coordinates are 11 32 49.92; +05 16 36.0. This star has no λ2175 absorption evident on either the *IUE l.d.* or *IUE h.d.* (high-dispersion) spectra. In the spectral region 1040Å-1060Å, the *FUSE* spectrum shows roughly equally intense B4R0, B4R1, and B4P1 lines, with the intensities of these three lines being about 3 times stronger than the other $H_2$ lines (B4R2, B4P2, B4R3, B4P3, B5P2, B5R3, and B5P3) present in the spectrum. No absorptions are seen at B5R4 or B5P4. The A I line at 1048.2Å is present, having about the same absorption intensity as the B4R0, B4R1, and B4P1 lines.

---

**#8  HD 149881**  (B0.5V) Coordinates are 16 36 58.20; +14 28 30.9. This star has no λ2175 on either *IUE l.d.* or *h.d.* The 1040Å-1060Å *FUSE* spectrum shows roughly equally strong B4R0, B4R1, and B4P1 lines, which appear to be 2-3 times wider than all the other $H_2$ lines in the spectrum, but this may partially be a result of saturation, since the former seem to be nearly totally absorbing at their peaks. The B4R2, B4P2, B4R3, B4P3, B5P2, B5R3, and B5P3 transitions are seen to absorb roughly 75% of the star's continuum light at their peaks. Absorption lines at B4R4, B5R4, and B5P4 are also seen in this star, with intensities 2-3 times less those of the transitions originating from J"=2 and J"=3. (The transition B4P4 at 1060.58Å falls just outside the *FUSE* spectral range here being monitored.) The A 1 absorption line is present, with an intensity comparable to the $H_2$ transitions originating from J"=2 and J"=3.  We regard star #8 as an example of a star that is just below threshold for stimulated Rayleigh scattering to occur.

---

**#9  HD 36541**  (B6V)  Coordinates are 05  32  7.03; -06  42  29.9.  This star has no λ2175 on *IUE l.d.* The *FUSE* 1040Å-1060Å spectrum closely resembles that of star #3, as far as absolute and relative intensities of $H_2$ lines are concerned. The A I line looks about the same in the two stars.

**#10  HD 37151** (B8V)  Coordinates are 05 36 6.23; -07 23 47.3.  This star has no λ2175 on *IUE l.d*. The *FUSE* 1040Å-1060Å spectrum is cluttered with many sharp - but relatively weak - absorption lines, which we have not tried to identify. It is hard to tell if there are any $H_2$ absorption lines present, but if any are, their intensities are rather weak. A sharp absorption line seen at a wavelength just slightly shorter than 1050Å may be B4R1 (rest frame wavelength: 1049.96Å). The A I line at 1048.2 Å is present.

______________________________________________________________________

**#11  HD 97991** (B1V)  Coordinates are 11 16 11.71; -03 28 19.1. This star has no λ2175 on *IUE h.d.* (a nice trace with relatively little noise). The *FUSE* 1040Å-1060Å spectrum looks generally similar to the *TUES* spectrum of star #6, as far as the relative observed intensities of the $H_2$ lines are concerned. However, in star #11 the lines originating from J"=2 and J"=3 are somewhat weaker, relative to those originating from J"=0 and 1, than in the case of star #6. This is perhaps reasonable, since star #11 is B1V, whereas star #6 is O9Vp.

______________________________________________________________________

**#12  HD 51013**  Coordinates are 06 54 41.24; -24 15 20.4. This star has no λ2175 on *IUE l.d.*  B4R1 is the only $H_2$ line appearing (very weakly) in the 1040Å-1060Å *FUSE* spectrum, which resembles that of star #1. HD 51013 was used as a reference star in ApJ **625,** 167 (2005) by Sofia *et al.*

______________________________________________________________________

**#13  BD-15 115  (**B2V)  Coordinates are 00 38 20.26; -14 59 54.2. This star shows no λ2175 absorption on *IUE l.d.* trace. The *FUSE* spectrum contains the A I line and many other sharp lines of medium intensity which we have not bothered to identify. If $H_2$ lines are present, they must be fairly weak.

______________________________________________________________________

**#14  GD712** (B3V)  Coordinates are 12 22 29.61; +40 49 35.6. This star shows no λ2175 absorption on *IUE l.d.* The *FUSE* spectrum is similar to that of star #11, only the $H_2$ lines are here a little bit stronger.

______________________________________________________________________

**#15  HD 3175** (B3V)  Coordinates are 00 34 22.39; -63 03 42.0. No λ2175 on *IUE l.d.* The *FUSE* spectrum is similar to that of star #13, with the Ar I line being the strongest absorber.

______________________________________________________________________

**#16  PB166** (B2V)  Coordinates are 13 24 0.72; +49 22 32.3. No λ2175 on nice *IUE l.d.* trace. The *FUSE* spectrum is similar to those in stars #13 and #15, with many non-$H_2$ lines, and with none of the lines being strong.

**#17 HD 205805** (B7III)  Coordinates are 21 39 10.61; -46 05 51.5. No λ2175 on nice *IUE l.d., h.d.* spectra. *FUSE* spectrum not unlike that of star #16, with perhaps there being even more lines here. If any are H$_2$ lines, they are weak.

___________________________________________________________________________

**#18 HD 45057** (B3V)  Coordinates are 06 22 18.84; -53 20 5.5. No λ2175 on *IUE l.d.* and *h.d.* spectra. The H$_2$ and A I line intensities in the *FUSE* spectrum are about as in star #4, but here there are additional absorption lines which we have not identified.

___________________________________________________________________________

**#19 HD 214080**  Coordinates are 22 36 6.44; -16 23 16.8. No λ2175 on *IUE h.d.* 1040Å-1060Å *TUES* spectrum has non-saturated linear H$_2$ spectrum.

___________________________________________________________________________

**#20 HD 214930**  Coordinates are 22 41 25.73; +23 50 47.7. No λ2175 on *IUE h.d.* No saturated H$_2$ lines on *TUES* spectrum. (*FUSE* spectrum not available.)

___________________________________________________________________________

**#21 HD 215733**  Coordinates are 22 47 2.51; +17 13 59.0. No λ2175 apparent on *IUE l.d.* or *h.d.* spectra. The *FUSE* spectrum is very much like that of star #8, only the H$_2$ lines are here somewhat more absorbing. Compared to star #8, star #21 must therefore be even closer to the threshold for stimulated Rayleigh scattering to occur.

___________________________________________________________________________

**#22 HD 42401**  Coordinates are 06 10 59.17; +11 59 41.5. In the 1040Å-1060Å *FUSE* spectrum of this star, the strengths of H$_2$ absorptions on transitions originating from levels J" =2,3, and 4 of X0 are about the same as in the case of star #21. However, in star #22 the relative strengths of H$_2$ lines originating from J"=0 and 1 (*i.e.* B4R0, B4R1, and B4P1) are considerably greater than in the case of star #21. In accord with this, medium-weak λ2175 absorption is seen to be present on the *IUE h.d.* trace of star #22. Star #22 is the first star thus far encountered in the present list that we would regard as definitely being "above threshold" for stimulated Rayleigh scattering to occur.

___________________________________________________________________________

**#23 AzV 377**  Coordinates are 01 05 7.33; -72 48 18.3. This star has no λ2175 on *IUE l.d.* and displays linear H$_2$ absorption in *FUSE* spectrum.

___________________________________________________________________________

**#24 AzV 372**  Coordinates are 01 04 55.73; -72 46 47.7. This star has no λ2175 on *IUE l.d.* and displays non-saturated linear H$_2$ absorption in the *FUSE* spectrum.

**#25 HD 186994** Coordinates are 19 45 37.97; +44 57 49.5. This star appears to have a small amount of λ2175 – barely visible on the *IUE l.d.* trace, but more clearly visible on the *IUE h.d.* trace. The strengths of the B4-0R0, B4-0R1, and B4-0P1 lines relative to those of B4-0R2, B4-0P2, B4-0R3, etc. in the *FUSE* spectrum are slightly greater than in the case of star #21, but are slightly less than in the case of star #22. Star #25 therefore appears to be the one "just above threshold" by the smallest margin among the twenty-five stars listed above.

---

**#26  BD+320270**  Coordinates are 01 34 50.39; +32 57 21.8. No apparent λ2175 on *IUE l.d.* trace. Pretty strong *linear* $H_2$ absorption on *FUSE* spectrum (plus many other lines). No sign of *nonlinear* H2 absorption.

---

**#27  HD 37332**  Coordinates are 05 37 45.89; -00 46 41.7. No apparent λ2175 on *IUE l.d.* No strong $H_2$ lines in *FUSE* spectrum.

---

**#28  HD 114444**  Coordinates are 13 13 4.15; -75 18 49.8. This star clearly has medium-weak λ2175 on *IUE l.d.*  In the *FUSE* spectrum, $H_2$ absorption appears more nonlinear than in star #25, but less nonlinear than in star #22.

---

**#29  HD 46150**  (O6V) Coordinates are 06 31 55.52; +04 56 34.3. Very strong λ2175 absorption on *IUE l.d.* and *h.d.* In the *FUSE* spectrum, the relative intensities of $H_2$ absorptions on transitions originating from (X0, J"=0) and (X0, J"=1) are huge compared to those on transitions originating from levels (X0, J"), with J" being greater than 1. See Fig. 4 of manuscript submitted to *Science*.

---

**#30  HD 217312** (B0 IV)  Coordinates are 22 58 39.80; +63 04 37.7. Very strong λ2175 on *IUE h.d.* The *FUSE* spectrum is very much like that of star #29, with saturated absorption occurring on $H_2$ transitions originating from (X0, J"=0 and 1), and with such absorption appearing to be at least a couple of orders of magnitude more intense than absorption on transitions originating from (X0, J"), with J" $\geq 2$.

---

**#31  HD 179406** (B3V)  Coordinates are 19 12 40.71; -07 56 22.3. Medium-strong λ2175 on *IUE l.d.* and *h.d.* The  (low noise) *FUSE* spectrum shows saturated $H_2$ absorption on transitions originating from (X0, J"=0 and 1), with such absorption being at least a couple of orders of magnitude more intense than absorption occurring on transitions originating from (X0, J") with J"$\geq 2$.

**#32 HD 41117** (B2 Ia) Coordinates are 06 03 55.19; +20 08 18.4. Very strong λ2175 on *IUE h.d.* and *l.d.* The (somewhat noisy) *FUSE* spectrum again shows strong, saturated $H_2$ absorption occurring on transitions originating from (X0, J"=0 and 1), while absorptions occurring on transitions originating from X0 levels with higher J" are at least a hundred times weaker.

---

**#33 HD 69106** (B0.5 IV) Coordinates are 08 14 03.84; -36 57 08.8. Medium-weak λ2175 on *IUE h.d.* and *l.d.* The *FUSE* $H_2$ spectrum is like that in star #22, but not quite as nonlinear as the latter.

---

**#34 HD 147888** (B3/B4 V) Coordinates are 16 25 24.3; -23 27 37.4. Strong λ2175 on *IUE l.d* and on a nice *WUPPE* trace. *FUSE* spectrum shows intensities of absorptions on transitions originating from (X0, J"=0 and 1) to be about 10 times those of transitions originating from (X0, J"=2), and at least 100 times those of transitions originating from (X0, J") levels with J" ≥3. Evidence of a P Cygni line profile on B4-0P2.

---

**#35 HD 46056** (B1V) Coordinates are 06 31 20.88; +04 50 3.3. Strong λ2175 on *IUE h.d.* and *l.d*. *FUSE* spectrum shows strongly saturated $H_2$ absorption on transitions originating from (X0, J"=0 and 1). These absorptions appear to be at least 100 times more intense than those of transitions originating from (X0, J") with J" greater than 1.

---

**#36 HD 97471** Coordinates are 11 12 7.00; -58 48 14.5. Medium-weak λ2175 on *IUE l.d*. The relative strengths and overall appearances of the $H_2$ absorptions in the *FUSE* spectrum are roughly the same as those seen in the spectra of stars #22 and #28.

---

**#37 HD 24263** (B5V) Coordinates are 03 52 0.23; +06 32 5.7. Clear medium λ2175 on *IUE h.d.* and also *l.d.,* with the former spectrum in this case being clearer. The 1040Å-1060Å *FUSE* spectrum shows greatly enhanced B4R0, B4R1, and B4P1, with B4R0 here appearing the most intensely absorbing. B4R2 and B4P2 are not enhanced.

---

**#38 HD 30122** (B5III) Coordinates are 04 45 42.47; +23 37 40.8. Has medium strength λ2175 on *IUE l.d.* (which here gets noisy at longer wavelengths). *FUSE* spectrum has saturated, greatly enhanced absorption at B4R0, B4R1, and B4P1.

**#39 HD 200775** (B2Ve) Coordinates are 21 01 36.92; +68 09 47.8. Wide, medium-strong λ2175 on *IUE l.d., h.d.* Very strong saturated absorption on B4-0R0, B4-0R1, B4-0P1, etc. Some enhancement clearly present on B4-0R2 and B4-0P2, etc. No enhancement on B4-0R3 and B4-0P3, etc.

---

**#40 HD 34078** (O9.5V) Coordinates are 05 16 18.15; +34 18 44.3. Strong λ2175 on both (IUE) h.d. and l.d. spectra. The FUSE spectrum shows strong, saturated, greatly enhanced absorption on B4R0, R1, and P1, etc. Some enhancement on B4R2 and P2, etc. Less enhancement on B4R3 and P3, etc.

---

**#41 HD 36665** (B0Ve) Coordinates are 05 34 39.10; +28 03 3.8. Strong λ2175 on IUE l.d. Very strong nonlinear $H_2$ absorption on B4R0, R1, P1 in FUSE spectrum. Some enhanced absorption also on B4R2 and P2.

---

**#42 HD 37318** (B1Ve) Coordinates are 05 38 58.02; +28 27 36.5. Very strong λ2175 on IUE h.d. Very strong nonlinear H2 absorption on R0, R1, and P1 lines originating from X0. All lines originating from higher J" values of X0 are not enhanced, except for R2 and P2, which are slightly enhanced.

---

**#43 HD 108** (O6 ps) Coordinates are 00 06 03.39; +63 40 46.8. Very strong λ2175 on IUE h.d. Very strong nonlinear saturated $H_2$ absorption on R0, R1, and P1. Some small enhancement on R2 and P2, about the same on R3 and P3. These four lines are all (equally) slightly broadened.

---

**#44 HD 123335** (B5 IV) Coordinates are 14 08 56.25; -59 16 36.2. Has only medium λ2175 on nice IUE l.d. The FUSE spectrum shows saturated R0, R1, and P1 lines, but these lines are much less strong than in stars #41-43, for example. Absorptions on transitions originating from X0 levels with J"≥2 are not enhanced or broadened.

---

**#45 HD 13621** (B1V) Coordinates are 02 14 32.97; +55 19 01.7. This star has medium-strong λ2175 on (somewhat noisy) IUE l.d. spectrum. Has saturated, enhanced R0, R1, and P1 absorptions which are somewhat stronger than in star #44, but less strong than in star #42.

**#46   HD 13854**  (B1 Iabe) Coordinates are 02 16 51.72; +57 03 18.9. This star shows broad, very strong λ2175 absorption in both its IUE l.d and h.d. spectra. The FUSE spectrum displays very strongly the broad, saturated $H_2$ absorptions on the B-X (R0, R1, P1) and C-X (R1, P1, Q1) transitions that always develop when the threshold for stimulated Rayleigh scattering is exceeded. The corresponding R2, P2, R3, and P3 absorption lines are slightly broadened in this star.

______________________________________________________________________

**#47   HD 14818**  (B2 Iae)  Coordinates are 02 25 16.03; +56 36 35.4. Very strong, wide λ2175 absorption appears on IUE h.d. and l.d. The FUSE spectrum shows the saturated $H_2$ absorption that canonically occurs on the B-X (R0, R1, and P1) and C-X (R0, R1, and Q1) transitions when the threshold for stimulated Rayleigh scattering is exceeded. P Cygni profile on absorption at B4P2.

______________________________________________________________________

**#48   HD 152245**  (B0 Ib)   Coordinates are 16 54 00.48; -40 31 58.1. Strong-medium λ2175 on IUE l.d. The FUSE spectrum again displays the saturated, greatly enhanced $H_2$ absorption that occurs on transitions originating from (X0, J"=0 and 1) when the threshold for nonlinear excitation (*i.e.* stimulated Rayleigh scattering) is exceeded. In this particular star no enhanced absorption is seen to occur on $H_2$ transitions originating from (X0, J"≥2).

______________________________________________________________________

**#49   HD 18352**  (B1V)   Coordinates are 02 59 47.37; +61 17 23.8.  Very strong λ2175 on IUE h.d. (especially) and also on IUE l.d. The FUSE spectrum shows the canonical saturated and enhanced absorption that occurs on $H_2$ transitions originating from (X0, J"=0 and 1) when the threshold for stimulated Rayleigh scattering is exceeded in the gaseous cloud enveloping the star. Some weak enhancement is seen on the lines originating from X0, J"=2. No enhanced absorption is here seen to occur on transitions originating from levels (X0, J"≥3).

______________________________________________________________________

**#50 HD 239729**  Coordinates are 21 39 27.40; +57 29 0.8. Strong, wide λ2175 on IUE l.d. Very strong, saturated, absorptions seen in the (relatively noisy) FUSE spectrum on $H_2$ transitions originating from (X0, J"=0 or 1).

______________________________________________________________________

**#51 HD 46149**  (O8.5V)  Coordinates are 06 31 52.53; +05 01 59.2. Very strong λ2175 on nice IUE l.d. and h.d. traces. The nice FUSE spectrum again displays the canonical nonlinear $H_2$ absorption spectrum, represented by overwhelming intensities of the B4R0, B4R1, and B4P1 absorptions in the 1040Å-1060Å range.

**#52 HD 148937** (O7V) Coordinates are 16 33 52.39; -48 06 40.5. Extremely strong λ2175 on IUE h.d. The FUSE spectrum is almost just like that of stars #51 and #29.

───────────────────────────────────────────────

**#53 HD 46223** (O4V) Coordinates are 06 32 9.31; +04 49 24.7. Extremely strong λ2175 on both the IUE h.d. and l.d. spectra. This star represents the hottest star on our list. Both the IUE and FUSE spectra of this O-type giant are remarkably similar to those of stars #29, #51, and #52. These four stars can therefore be viewed as perfect examples of O-type giants surrounded by gaseous clouds containing high enough densities of $H_2$ molecules that the latter are way above threshold for stimulated Rayleigh scattering to occur.

───────────────────────────────────────────────

**#54 HD 192639** (O7 I) Coordinates are 20 14 30.43; +37 21 13.8. Another O-type star (in this case a supergiant) which displays extremely strong λ2175 absorption in its UV spectrum. (The IUE h.d. spectrum is especially impressive.) The FUSE spectrum shows again the canonical spectrum of a star in which $H_2$ molecules in the gaseous cloud surrounding the star are being very strongly nonlinearly driven.

───────────────────────────────────────────────

**#55 HD 74194** (O8.5 Ib) Coordinates are 08 40 47.79; -45 03 30.2. Another O-type supergiant with very strong λ2175 absorption on IUE h.d. The *FUSE* spectrum of this star is similar to that of star #54, but is slightly less nonlinearly absorbing than the latter, as would be expected from the fact that this star is slightly cooler (O8.5 vs. O7).

───────────────────────────────────────────────

**#56 HD 157857** (O6.5IIIf) Coordinates are 17 26 17.33; -10 59 34.8. Very strong λ2175 absorption on IUE l.d. The *FUSE* spectrum looks very much like that of star #52, for example.

───────────────────────────────────────────────

**#57 HD 220057** (B2 IV) Coordinates are 23 20 0.65; +61 08 59.4. Medium-weak λ2175 on IUE l.d. The FUSE spectrum shows saturated absorption at B4R0, R1, and P1, but the widths of the spectral regions that are totally saturated are somewhat less than in the case of, say, star #55.

───────────────────────────────────────────────

**#58 HD 698** (B5 II) Coordinates are 00 11 37.15; +58 12 42.6. Medium, but very broad, λ2175 on both IUE l.d. and h.d. The widths of the spectral regions over which the B4-0R0, R1, and P1 absorptions show complete saturation are somewhat greater than in the case of, say star #57.

**#59 HD 9234** (B8…)  Coordinates are 01 32 5.48; +54 01 8.3. Medium strength λ2175 absorption on (somewhat noisy) IUE l.d. spectrum. The *FUSE* spectrum shows saturated $H_2$ nonlinear absorptions centered at B4-0R0, R1, and P1. These have about the same general appearance as those in star #58. Absorptions on $H_2$ transitions originating from (X0, J"=2) are here slightly broadened. Absorptions on transitions originating from (X0, J"=3) are broadened still less. No broadening of absorptions on transitions originating from still higher rotational levels of X0 can be clearly discerned.

___________________________________________________________________________

**#60 HD 216014** (B0.5V)  Coordinates are 22 47 52.94; +65 03 43.8. Strong-to-very-strong λ2175 absorption observed on IUE l.d.  The FUSE spectrum is quite like that of star # 51, for example. As we continue to examine the stars in this list, it really seems more and more plausible that such a spectrum does indeed represent a star surrounded by a gaseous cloud in which $H_2$ molecules are undergoing stimulated Rayleigh scattering.

___________________________________________________________________________

**#61 HD 198478** (B3 Ia)  Coordinates are 20 48 56.29; +46 06 50.9. Strong-to-very strong λ2175 on IUE l.d., h.d.  The FUSE spectrum shows strongly saturated enhanced absorption on B4R0, B4R1, and B4P1. There is some broadening of absorption on B4P2. Absorptions on transitions originating from rotational levels of X0 higher than J"=2 are not significantly broadened. The FUSE spectrum appears to contain some P Cygni –like profiles.

___________________________________________________________________________

**#62 HD 21641** (B8.5V)  Coordinates are 03 31 33.13; +47 51 44.7. This seems to be a "Just Over Threshold (JOT)" star. The IUE l.d. spectrum shows weak, broad λ2175 absorption. The FUSE spectrum shows enhanced absorption on B4R0, B4R1, and B4P1, but, although these absorptions are all saturated at their peaks, the spectral widths over which saturated absorption prevails are much less than in any of the stars for which the λ2175 absorption is more pronounced. No broadening or enhancement of any kind is evident in the absorptions on transitions originating from rotational levels of X0 higher than J"=1.

___________________________________________________________________________

**#63 HD 20683** (B9V)  Coordinates are 03 23 47.33; +48 36 15.9. Compared to star #62, this star seems to have even less λ2175 absorption apparent in the IUE l.d. spectrum. The FUSE spectrum of this star is very similar to that of star #62. Note that these two stars have almost the same coordinates.

**#64　HD 21279**　(B8.5V)　Coordinates are 03 27 55.78; +47 44 9.4. This star has weak, broad λ2175 absorption evident in the IUE l.d. spectrum. The FUSE spectrum shows saturated absorptions at B4R0, B4R1, and B4P1 that are somewhat wider and more strongly enhanced with respect to B4R2, B4P2, B4R3, B4P3, etc. than in the cases of stars #62 and #63.

───────────────────────────────────────────────────────────────

**#65　HD 212791**　(B8…)　Coordinates are 22 25 41.75; +52 26 18.4. This star shows a definite (but small) amount of λ2175 in its IUE l.d. spectrum. The FUSE spectrum is very much like that of star #21.

───────────────────────────────────────────────────────────────

**#66　HD 21551**　(B8V)　Coordinates are 03 30 36.95; +48 06 12.9. Another star that is either a "Just Over Threshold (JOT)" or "Just Under Threshold (JUT)" star. (This is difficult to discern from the IUE l.d. spectrum.) The FUSE spectrum here is generally similar to that of star #22, but the enhancement of the absorptions occurring on B4R0, R1, and P1 (relative to those occurring about the other $H_2$ transitions in the 1040Å-1060Å spectrum) are here less.

───────────────────────────────────────────────────────────────

**#67　HD 21672**　(B8V)　Coordinates are 03 31 53.95; +48 44 6.5. All comments made above about the IUE l.d. and FUSE spectra of star #66 apply also in the case of this star.

───────────────────────────────────────────────────────────────

**#68　HD 163522**　(B1 Ia)　Coordinates are 17 58 35.23; -42 29 10.1. This star has definite medium-weak λ2175 on IUE h.d. The FUSE spectrum shows saturated B4R0, B4R1, and B4P1 absorptions, with the intensity enhancements being more than those seen in star #21 and less than those seen in star #22.

───────────────────────────────────────────────────────────────

**#69　HD 22136**　(B8V)　Coordinates are 03 35 58.49; +47 05 27.7. This star appears to have broad, weak λ2175 absorption on the IUE l.d. spectrum. In the FUSE spectrum, saturated absorptions about the B4R0, R1, and P1 transitions are seen, with intensities enhanced over those of absorptions on transitions originating from higher-J" levels of X0 by about the same amount as in the case of star #22.

**#70  HD 216898**  Coordinates are 22 55 42.46: +62 18 22.8. This is a star that is very strongly nonlinearly excited. This star has extremely strong λ2175 absorption apparent on both its IUE l.d. spectrum, and also on a. very low noise WUPPE spectrum. The FUSE spectrum shows the canonical signature of $H_2$ in the gaseous cloud surrounding the star being very strongly nonlinearly photoexcited via stimulated Rayleigh scattering. One sees all the light from the star being totally absorbed over a wide spectral region that completely encompasses the wavelengths of the B4R0, B4R1, and B4P1 transitions. There is some, but orders of magnitude less, enhanced absorption occurring about the B5P2, B4R2, and B4P2 transitions, and even a very slight amount of enhancement occurring on the transitions that originate from (X0, J"=3)

―――――――――――――――――――――――――――――――――――――――――――――――――――――――――

**#71  HD 216532**  Coordinates are 22 52 30.56; +62 26 26.0. Extremely strong λ2175 on IUE l.d. Same comments about the FUSE spectrum of star #70 apply in the case of this star, although the signal-to-noise ratio is here not quite as good.

―――――――――――――――――――――――――――――――――――――――――――――――――――――――――

**#72  HD 217086**  Coordinates are 22 56 47.19; +62 43 37.6. All comments made for star #71 also apply here.

―――――――――――――――――――――――――――――――――――――――――――――――――――――――――

**#73  HD 219188**  Coordinates are 23 14 0.57; +04 59 49.5. This is either a JOT or JUT star. The IUE h.d. spectrum seems to indicate that a weak, broad λ2175 absorption is present. The FUSE spectrum is generally similar to that of star #21.

―――――――――――――――――――――――――――――――――――――――――――――――――――――――――

**#74  HD 206165**  (B2 Ib)  Coordinates are 21 37 55.22; +62 04 55.0. Strong λ2175 present on both IUE h.d. and l.d. The FUSE spectrum is very much like that of star # 61.

―――――――――――――――――――――――――――――――――――――――――――――――――――――――――

**#75    HD 37903**    Coordinates are 05 41 38.39. −02 15 32.5. Medium-strong λ2175 absorption is clearly evident on both the IUE l.d. and h.d. spectra, and is also present on a WUPPE spectrum.  The FUSE spectrum shows strong, saturated, enhanced absorption present about the B4R0, R1, and P1 transitions. What is unusual about this star is that the absorptions at B5P2, B4R2, and B4P2 are both stronger and wider than usual. The R3 and P3 absorptions are also slightly enhanced.

―――――――――――――――――――――――――――――――――――――――――――――――――――――――――

**#76  HD 122879**  (B0 Ia)  Coordinates are 14 06 25.16; -59 42 57.3. Strong λ2175 clearly present on nice IUE l.d. and h.d. spectra.  The FUSE spectrum is very much like that of star #48 - not surprising in view of the fact that the latter is a B0 Ib star.

**#77  HD 37367**  Coordinates are 05 39 18.31; +29 12 54.8. Very strong λ2175 on IUE l.d. The FUSE 1040Å-1060Å spectrum displays the canonical saturated and strongly enhanced absorptions around B4R0, B4R1, and B4P1 that we associate with $H_2$ being nonlinearly photoexcited in the gaseous cloud surrounding the star. Besides absorptions occurring around $H_2$ transitions, there are many narrow absorptions which we have not identified in this spectrum.

---

**#78  HD 14250**  Coordinates are 02 20 15.73; +57 05 55.0.  Strong λ2175 absorption on IUE l.d. seen in this star. The (somewhat noisy) FUSE spectrum shows the canonical wide spectral region of saturated absorption that spans the B4R0, R1, and P1 transitions when the $H_2$ in the gaseous cloud surrounding the star is nonlinearly photoexcited.

---

**#79  HD 43384**  (B3 Iab)  Coordinates are 06 16 58.71; +23 44 27.3. Very strong λ2175 on (somewhat noisy) IUE h.d. The (also somewhat noisy) FUSE spectrum shows the canonical wide spectral region of saturated absorption spanning the B4R0, R1, and P1 transitions that occurs when $H_2$ in the gaseous cloud surrounding the star is nonlinearly photoexcited via the mechanism of stimulated Rayleigh scattering. The FUSE spectrum of this star looks almost identical to that of star #61 – not surprising, in view of the fact that the spectral type of the latter star is also (B3 Ia). (Note that the coordinates of the two stars are entirely different.) In the FUSE spectra of both stars, the absorptions that occur about $H_2$ transitions that originate from (X0, J''=2) and (X0, J''=3) appear to have definite P Cygni profiles, signifying that $H_2$ molecules in these levels also undergo stimulated Rayleigh scattering to some extent.

---

**#80  HD 190603**  (B1.5 Ia)  Coordinates are 20 04 36.17; +32 13 7.0. Very strong λ2175 on both IUE l.d. and h.d. The FUSE spectrum again shows the signature of strong nonlinear $H_2$ photoexcitation occurring in the cloud surrounding the star. In many ways the FUSE spectrum of this star is similar to that of star #79, with some P Cygni profiles being present, etc. However, somewhat paradoxically, the widths of the spectral regions over which the absorption is totally saturated in the case of this star are slightly less than in star #79, which is a cooler supergiant.

---

**#81  HD 92964**  (B2.5 Iae)  Coordinates are 10 42 40.57; -59 12 56.7. Again one has here a hot supergiant with strong λ2175 occurring on IUE h.d. and l.d. The FUSE spectrum looks almost identical with that of star #61.

**#82  HD 14143**  Coordinates are 02 19 13.89; +57 10 9.8. Very strong λ2175 on both IUE h.d. and l.d. The (somewhat noisy) FUSE spectrum shows the canonical heavily saturated regions of nonlinear absorption about all $H_2$ transitions originating from (X0, J"=0 or 1).

---

**#83  HD 14134**  Coordinates are 02 19 4.45; +57 08 7.8. Very strong λ2175 on IUE l.d. The (somewhat noisy) FUSE spectrum looks almost identical to that of star #82, which has very similar coordinates. Both stars display clear P Cygni profiles on the B4P2 transition.

---

**#84  BD+56 510**  Coordinates are 02 18 54.27; +57 09 29.9. Strong-to-very strong λ2175 is seen on the IUE l.d. spectrum of this star, which has coordinates very close to those of stars #82 and #83. The (1040Å-1060Å) FUSE spectrum of this star again shows greatly enhanced absorption about B4R0, B4R1, and B4P1. In this star, the total spectral width over which the nonlinear absorption about the above three transitions is totally saturated is about 300 wavenumbers - a typical value for most of the "above threshold" stars in the present list. The P Cygni profiles about B4P2 and B5P2 are much less prominent in star #84, compared to those in star #83.

---

**#85  HD 21291**  (B9 Ia)  Coordinates are 03 29 4.13; +59 56 25.2. This star does have very strong λ2175 on IUE h.d. The (somewhat noisy) FUSE spectrum displays the canonical 300-wavenumbers-wide region of saturated absorption about B4R0, B4R1, and B4P1 that (we believe) signifies the occurrence of $H_2$ nonlinear photoexcitation (*i.e.* stimulated Rayleigh scattering) in the gaseous cloud surrounding the star.

---

**#86  HD 225094**  (B3 Ia)  Coordinates are 00 03 25.71; +63 38 25.9. This again is a star with the same spectral type as star #61. It is therefore no surprise that everything about the two stars is the same. Both stars show very strong λ2175 on their IUE h.d. and l.d. spectra. Furthermore, the FUSE spectra of these two stars are virtually identical. Therefore, the comments that were made regarding evidence of nonlinear $H_2$ photoexcitation in star #61 also fully apply here.

---

**#87  HD 53367**  Coordinates are 07 04 25.53; -10 27 15.7. This star has strong-to-very strong λ2175 absorption on IUE h.d and l.d. The FUSE spectrum shows the canonical 300-wavenumbers-wide region of totally saturated absorption occurring about B4R0, R1, and P1. In this spectrum, the absorptions at B5P2 and B4P2 appear broadened to a greater degree than is observed in the spectra of virtually all the other stars in this list. However, absorptions at all the other $H_2$ transitions which one normally sees between 1040Å and 1060Å (*i.e.* B5R3, B5P3 B5R4, B5P4, B4R3, B4P3, and B4R4) are not detectably broadened.

**#88 HD 47129** (O8V + O8f)  Coordinates are 06 37 24.04; +06 08 7.4. This star has medium-to-strong λ2175 absorption clearly evident on nice IUE l.d. and h.d. spectra. The (low-noise) FUSE spectrum again perfectly displays the canonical signature of $H_2$ being strongly nonlinearly photoexcited in the gaseous cloud surrounding the star. The B4R0, R1, and P1 absorptions in the spectrum are enormously enhanced, while the absorptions about all other $H_2$ transitions in the spectrum are not.

---

**#89 HD 46202**  Coordinates 06 32 10.47; +04 57 59.8. Strong λ2175 absorption on IUE h.d. The FUSE spectrum shows the canonical saturated and enormously enhanced absorptions on transitions originating from (X0, J"=0 and 1). Very little or no enhanced absorption about transitions originating from higher-J" levels of X0.

---

**#90 HD 111973** (B5 Ia)  Coordinates are 12 53 48.92; -60 22 34.5. Strong λ2175 absorption on IUE l.d. The FUSE spectrum does show greatly enhanced absorption occurring about B4R0, R1, and P1. However, the combined spectral widths of the fully saturated regions around B4-0 in the case of this star is only about 150 wavenumbers, rather than the usual value of about 300 wavenumbers. There are some P Cygni profiles on some of the transitions originating from J"=2 and J"=3 of X0.

---

**#91 HD 111990**  Coordinates are 12 53 59.80; -60 20 7.5. Strong-to-very strong λ2175 absorption on IUE l.d. The FUSE spectrum is like that of star #90, except that in star #91 the combined spectral widths of the fully saturated regions around B4-0 is somewhat greater.

---

**#92 CPD-59 4552**  Coordinates are 12 53 46.47; -60 24 12.3. Basically all that was noted above for star #91 also applies in the case of this star. The FUSE spectra are very similar, and the strengths of the λ2175 absorptions as seen in the IUE l.d. spectra are about the same.

---

**#93 HD 42087** (B2.5 Ibe)  Coordinates are 06 09 43.98; +23 06 48.5. It is very telling that the profiles of the (strong) λ2175 absorption band seen in the IUE h.d. and l.d. spectra of this star are observed to be completely like those seen in the IUE h.d. and l.d. spectra of star #81, which also happens to be a B2.5 supergiant. Additionally, the FUSE spectra of the two stars are seen to be virtually the same. Adding further significance to the above observations is the fact that the coordinates of the two stars are entirely different.

**#94  HD 91983**   Coordinates are 10 35 54.18; -58 15 27.4. This star shows medium intensity λ2175 in the IUE l.d. spectrum. The FUSE spectrum shows absorption around B4-0R0, R1, and P1 being enhanced with respect to absorption on the transitions originating from higher J" values of X0 somewhat more than in the case of star #22.

---

**#95 HD 91943**  Coordinates are 10 35 42.01; -58 11 34.4. Weak, broad λ2175 on IUE l.d. This "Just Over Threshold (JOT)" star has a FUSE spectrum looking like that of star #25.

---

**#96  HD 73882**  Coordinates are 08 39 9.53; -40 25 9.3. This star shows medium-to-strong λ2175 on IUE l.d. and h.d., as well as on a noise-free WUPPE spectrum. The FUSE spectrum shows complete saturated absorption occurring in a 400-wavenumbers-wide spectral interval encompassing the B4R0, R1, and P1 transitions. Some degree of enhancement is also seen on the B5P2, B5R3, B5P3, B4P2, B4R3, and B4R4 transitions, with the P2 transitions being the most enhanced in this latter group.

---

**#97 HD 197512**  Coordinates are 20 42 10.05; +49 44 5.1. Medium-strong λ2175 on IUE l.d. The FUSE spectrum shows strongly-enhanced, saturated absorption about B4R0, R1, and P1.

---

**#98   HD 210121**  Coordinates are 22 08 11.90; -03 31 52.8. Medium λ2175 seen on (somewhat noisy) IUE l.d. and h.d. spectra. The (somewhat noisy) FUSE spectrum shows saturated absorption occurring in a 300-wavenumbers-wide spectral interval spanning the B4R0, R1, and P1 transitions. The R2 and P2 transitions are slightly enhanced. No enhancement is seen to occur on transitions originating from higher rotational levels of X0.

---

**#99  HD 113012**  Coordinates are 13 01 45.53; -60 04 35.2. This star has medium strength λ2175 on a (somewhat noisy) IUE l.d. spectrum. In the FUSE spectrum, nonlinear $H_2$ absorption is somewhat more enhanced than in star #94.

---

**#100  BD +56 524**  Coordinates are 02 19 6.44; +57 07 35.0. Strong λ2175 on noisy IUE l.d. spectrum. The (also noisy) FUSE spectrum shows strong nonlinear $H_2$ absorption around B4R0, B4R1, and B4P1.

---

**#101  CD-57 3348**  Coordinates are 10 35 46.57; -58 14 12.1. Weak, broad λ2175 on noisy IUE l.d. The FUSE spectrum shows somewhat less enhanced absorption occurring at the B4R0, R1, and P1 transitions than occurs in the case of star #94.

**#102   HD 116852**   Coordinates are 13 30 23.52; -78 51 20.5. Weak, broad $\lambda 2175$ definitely present on nice IUE l.d. (also on IUE h.d.) spectrum. The nice FUSE spectrum shows that absorptions occurring on the B4R0, R1, P1 transitions are enhanced over absorptions occurring on all other $H_2$ transitions in the 1040Å-1060Å range by roughly the same amount as in star #22.

___

**#103   HD 99872**   Coordinates are 11 28 18.41; -72 28 26.3. This star has medium-strong $\lambda 2175$ on nice IUE l.d. spectrum. The FUSE spectrum shows greatly enhanced B4R0, R1, and P1.

___

**#104   HD 35899**   (B5V)   Coordinates are 05 27 0.19; -21 18 43.4.  No $\lambda 2175$ on IUE l.d. spectrum. The FUSE spectrum shows only linear $H_2$ absorption, with absorption at B4R0, R1, and P1 not being enormously greater than that at other $H_2$ transitions occurring in the 1040Å-1060Å range.

___

**#105   HD 197770**   (B2III)   Coordinates are 20 43 13.68; +57 06 50.4. Very strong $\lambda 2175$ on IUE l.d., h.d., and on WUPPE spectra. The FUSE spectrum shows strongly enhanced, completely saturated absorption about B4R0, R1, and P1.